# Comparison calibration system for digital-output infrasound sensors

Koto Hirano[1,2]*, Hironobu Takahashi[1], Keisuke Yamada[1], Hideaki Nozato[1], and Shuichi Sakamoto[2]
[1] National Metrology Institute of Japan, National Institute of Advanced Industrial Science and Technology, 1-1-1 Umezono, Tsukuba, Ibaraki 305-8563, Japan
[2] Research Institute of Electrical Communication / Graduate School of Information Sciences, Tohoku University 2-1-1, Katahira, Aoba-ku, Sendai-shi, Miyagi, 980-8577 Japan

* Author to whom any correspondence should be addressed.
**Email**: koto.hirano@aist.go.jp

## Abstract
Recent advancements in micro electro-mechanical systems (MEMS) have enabled the application of digital-output MEMS modules in infrasound monitoring. These modules, combining MEMS pressure sensors and microcontrollers, provide timestamped digital pressure data. Compared with conventional analog infrasound sensors, the affordability and compactness of MEMS modules allow the construction of infrasound monitoring networks with a high density of measurement stations. However, dynamic frequency response characteristics of the MEMS modules, including both sensitivity modulus and phase, remain unassessed. In this study, we developed a comparison calibration system for digital-output infrasound sensors, with special attention paid to ensuring phase synchronization between the analog-output of reference standards and the digital-output of MEMS modules. Using a pulse per second signal synchronized with a time frequency standard, we successfully timestamped the reference analog signals, achieving synchronization between reference analog standards and digital-output sensors. Example calibrations were conducted on a digital-output MEMS module consisting of a DPS310 MEMS pressure sensor and an ESP32 microcontroller, in the 0.2 Hz to 4 Hz range. The sensitivity modulus matched the reference within a few percent, but the phase delayed by approximately 10 ms. We anticipate that the appliying corrections based on the results reported herein will enhance the reliability of infrasound measurements with digital-output sensors.

---

# 1 Introduction

Infrasound measurements play a crucial role in the monitoring of large-scale natural disasters, such as volcanic eruptions and tsunamis, as well as anthropogenic events, such as nuclear tests [1]. To capture infrasound signals, various organizations, including academic institutions, governmental agencies and international bodies, have been independently developing and operating infrasound networks of various sizes [2-4]. As a well-known worldwide network, the International Monitoring System (IMS), established under the Preparatory Commission for the Comprehensive Nuclear-Test-Ban Treaty Organization (CTBTO), operates approximately 60 stations worldwide to detect possible nuclear tests [2].

Typically, an infrasound network is composed of multiple infrasound sensors installed at different stations. The signals observed at each station are gathered and analysed based on their specific purposes. For example, in sound source localization, the source location is estimated based on the phase differences between multiple infrasound waveforms recorded at different stations. In such cases, the number of observation stations is a key factor that directly affects the performance of an infrasound network. However, from a practical standpoint, installing a large number of infrasound sensors is not feasible because of cost limitations: conventional sensors, such as analog microphones and barometers, are accurate but expensive [5-7].

Recently, the advancement of micro electro-mechanical systems (MEMS) has expanded the use of MEMS-based digital-output measurement modules to infrasound monitoring [8, 9]. These MEMS modules consist of two main parts: a MEMS pressure sensor that measures pressure, and a microcontroller that controls the sensor. The affordability of these modules compared to analog sensors enables their extensive deployment. In addition, digital pressure data from MEMS pressure sensors are timestamped by the microcomputer, facilitating the easy comparison of infrasound waveforms recorded at different locations.

Although the use of digital-output MEMS modules has been expanding, almost all manufacturers provide specifications only for static pressure characteristics of the MEMS pressure sensors, because these are primarily intended for static pressure measurements. Their dynamic characteristics, such as frequency response—including sensitivity modulus and phase—have not been comprehensively assessed. Notably, the calibration of digital-output infrasound sensors, particularly phase calibration, is challenging. The reason is twofold: (1) calibration in the infrasonic range is still being developed and (2) calibration of digital-output sensors requires different handling techniques, such as timing synchronization, compared to analog sensors, which have traditionally been the focus of acoustic calibrations. As for reason (1), primary calibration methods for frequencies below a few Hz, where traceable calibration techniques are yet to be established, have been extensively developed by National Metrology Institutes and research organizations [10-16]. Furthermore, comparison calibrations of field-deployed instruments using primary-calibrated instruments as reference standards are currently under development [17-24]. Nevertheless, at present, comparison calibration mainly focuses on high-precision analog microphones and barometers, such as SeismoWave MB3a and Chaparral Physics Model 25 used in the CTBT monitoring network [5, 6]. To the best of our knowledge, sensitivity modulus and phase calibration of digital-output MEMS modules have not yet been reported. We attribute this to reason (2): MEMS modules are usually calibrated through comparison because their size and other parameters are not suitable for primary calibration. When comparing them with analog reference sensors, special attention is required to ensure phase synchronization between the analog-output of the reference and the digital-output of the MEMS modules.

In the present study, to calibrate digital-output MEMS modules, we first considered the general requirements for the calibration of digital-output sensors. Next, we developed a comparison calibration system that meets these requirements. Considering phase synchronization, we constructed a system that applies timestamps to the analog reference signals using a pulse per second (PPS) signal synchronized with the time standard generated by the National Metrology Institute of Japan, namely UTC(NMIJ). Using this system, we calibrated the sensitivity modulus and phase of MEMS modules, consisting of a MEMS pressure sensor (DPS310) and a microcontroller (ESP32) [25, 26].

The remainder of this paper is structured as follows: Section 2 describes specifications for comparison calibration systems applicable to digital-output infrasound sensors. Section 3 provides experimental details. Section 4 presents the calibration results. In Section 5, uncertainty components are discussed. Finally, Section 6 concludes this study.

# 2 Calibration system: requirements and implementation

## 2.1 General requirements

The key points for a comparison calibration system that can be applied to digital-output sensors are summarized below:

**(1)** The comparison calibration system should expose both the device under test (DUT) and the reference standard to the same sound pressure over the target frequency range. This is relatively easy in the low frequency range, where the sound wavelengths are on the order of meters, which are substantially larger than the dimensions of the devices.

**(2)** Such a calibration system must ensure that the analog-output from the reference standard and the digital-output from the DUT are synchronized. This requirement is particularly important for the calibration of digital-output sensors. In general acoustic calibration, where the DUTs are analog sensors such as microphones, the analog voltages from the reference microphone and the DUT microphone are digitized using the same digitizer. As long as synchronization between the channels of the digitizer is ensured, their waveforms, including phase information, are easy to compare. However, when evaluating a digital-output DUT, the outputs of the DUT are timestamped ASCII or binary data whereas the reference microphone outputs an analog voltage proportional to the sound pressure. Therefore, to compare their waveforms, either requirement A or B should be fulfilled:

**A)** The start time of analog signal acquisition must be precisely synchronized with that of the digital DUT.

**B)** An accurate timestamp must be assigned when digitizing the reference analog signal, so that the output of the reference and the DUT can be compared on the same absolute time axis.

## 2.2 Design strategies for fulfilling the requirements

Herein, we explain the design strategies employed to ensure that the calibration system has characteristics (1) and (2) in Section 2.1.

**(1)** We designed chamber dimensions to ensure sufficient space for both the reference standard and the DUT, while maintaining uniform sound pressure in the infrasound range, i.e. below 20 Hz. The internal dimensions were designed to be less than one twentieth of the wavelength, specifically 220 mm × 300 mm × 250 mm. It was confirmed that in this range, the uniformity of sound pressure is maintained within ±0.6% for modulus and ±0.1° for phase. The chamber was equipped with a sealed loudspeaker capable of generating sinusoidal sound pressure waves from 0.2 Hz to 20 Hz, with amplitudes up to 2 Pa.

**(2)** We developed a signal-acquiring system to satisfy requirement (B), considering that fulfilling requirement (A) is difficult because the time required for device communication is not constant. The basic scheme is presented in Figure 1(a). The key idea is that an analog/digital (A/D) converter was instructed to simultaneously sample the reference analog voltage $v_\mathrm{a}$ and the PPS voltage signal $v_\mathrm{b}$ (Figure 1(b)). The timestamp to start acquiring the reference analog signal was obtained in two steps. First, the fractional-second-only start time was determined by monitoring the PPS signal whose rising edge corresponds to the exact start of each second. Second, the coarse start time (accurate to the nearest second or better) was obtained by recording the system time on the PC at the moment it sends the start command to the A/D converter. For example, if the time recorded by the PC for the start command is (hh.mm.ss.???), and the A/D converter detects the rising edge of the PPS signal (??.??.??.000) after $t_p$ (ms), the absolute start time of the A/D conversion can be estimated as (hh.mm.ss.(1000−$t_p$)), assuming the time recorded by PC is accurate within 1 second.

Independently, the MEMS module was powered on, the system clock of the microcontroller was reset, and recording of pressure data $p_d$ with timestamps in the format of ($hh_d.mm_d.ss_d.SSS_d$) began just before the analog system started measurement. Later, the data group with timestamps close to those of the analog measurements was extracted for analysis.

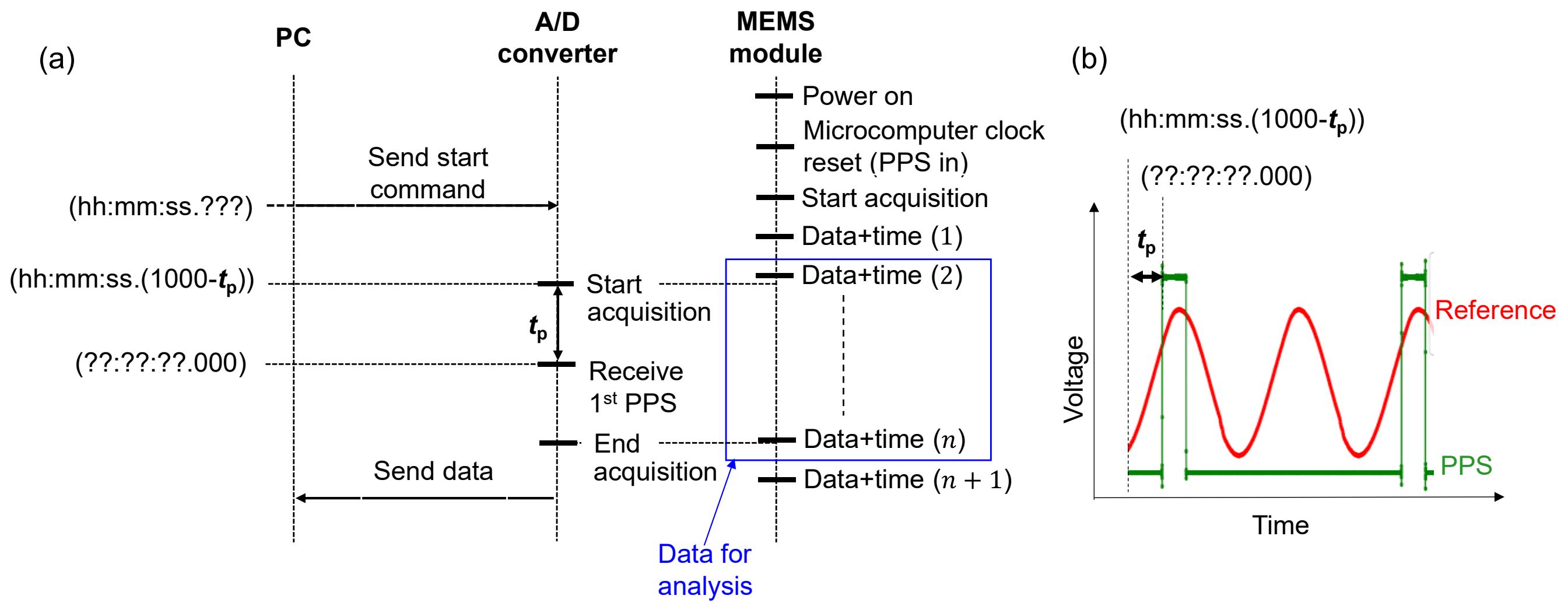

Figure 1. Scheme for assigning the timestamp to the reference analog signal. (a) Communication sequence between the PC and the A/D converter, and measurement process using the MEMS module. (b) Timing relationship between the reference analog signal and the PPS signal.

# 3 Experiment

## 3.1 Setup details

The experimental setup is shown in Figure 2. The sine wave was generated by the National Instruments (NI) PXIe-4461 D/A converter, amplified by the Luxman B-1 power amplifier, and then fed into a loudspeaker, yielding approximately 2 Pa of sound pressure inside the chamber. Then, the outputs of the reference standard and DUT were compared. For the reference standard, a Brüel & Kjær (BK) Type 4160 laboratory standard microphone (S/N: 2583072), whose pressure sensitivity was calibrated using a laser pistonphone method, was employed [27, 28]. Only the microphone diaphragm was exposed to the sound field, corresponding to the pressure-field measurement. To determine the exact start time of analog data acquisition, both the microphone voltage and the PPS signal were sampled, as mentioned in Section 2.2. Although using the same A/D converter for both signals would be ideal, separate converters were selected to better suit the specific requirements for each signal. The PPS signal, with its narrow pulse width (20 μs), required a high sampling rate, while the microphone output benefited from a high-resolution converter. Therefore, the NI PXIe-5172 was used for the PPS signal, and the PXIe-4461 A/D converter was used for the microphone. Synchronization between the digitizers was achieved by synchronizing their on-board clocks to a 10 MHz frequency-reference signal from a FS725 rubidium frequency standard. The sampling frequency for the microphone voltage was set to more than 1,000 times the target frequency, whereas the PPS signal was sampled at a frequency of at least 500 kHz. The PPS signal from the Freqtime time-distributer was synchronized to UTC(NMIJ).

Independently, the data from the MEMS module was downloaded from the external network time protocol (NTP) server. The details for data acquisition are explained in Section 3.2. The frequency response of the MEMS module was evaluated by comparing the data from the microphone with those from the MEMS module (Note that this refers to the overall response of the module, not the individual response of the MEMS pressure sensor or the microcontroller).

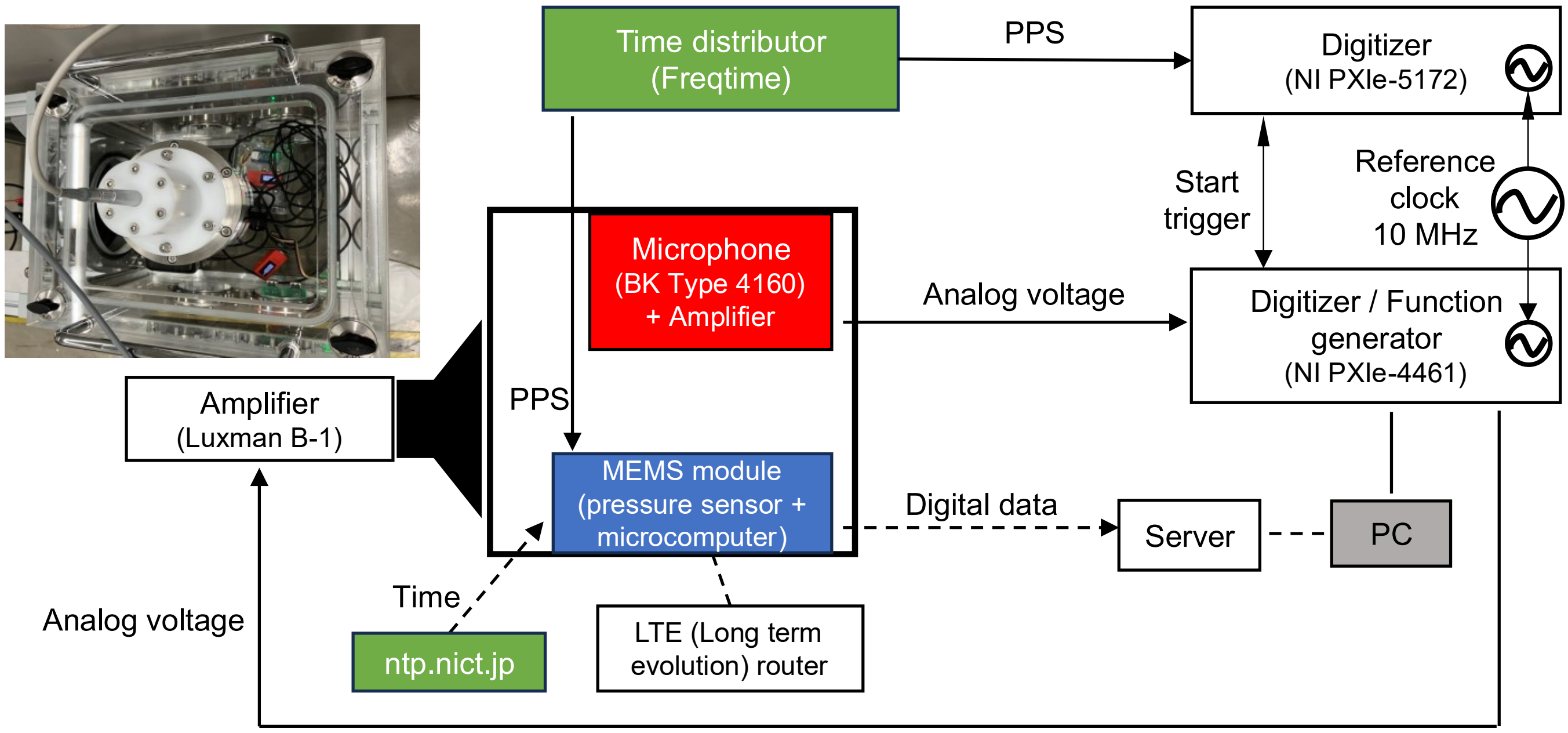

Figure 2. Schematic of the experimental setup. The dotted line and the solid line represent wireless communication and wired communication, respectively.

## 3.2 MEMS module evaluated in this study

To validate the proposed comparison calibration system, we evaluated MEMS modules which composed of a MEMS pressure sensor (DPS310) responsible for pressure measurement and a microcontroller (ESP32) which controls the pressure sensor and attaches timestamp information to the pressure data (Figure 3) [25, 26]. The roles of each component are described below. Upon receiving a measurement command from the controller, the MEMS pressure sensor measures the atmospheric pressure and transmits the measured value (or the corresponding numerical value). Since the MEMS pressure sensor itself does not have time-related information, the controller obtains the time from an external time-source and attaches the timestamp $t_{\mathrm{d}}$ immediately after receiving the pressure value $p_{\mathrm{d}}$, thereby associating each pressure value with corresponding time. To enhance the accuracy of timestamp acquisition, we improved the time acquisition system as described in the Appendix.

Pressure fluctuation i.e. sound pressure can be measured by repeatedly measuring pressure at fixed intervals (in this case 70 ms) and monitoring its changes. Once a sufficient amount of data has been accumulated, the microcontroller uploads the pairs of $(t_{\mathrm{d}}, p_{\mathrm{d}})$ to an external cloud server.

It should be noted that the MEMS module evaluated in this study is only one example. The proposed approach is also applicable to other MEMS modules composed of different microcontrollers and MEMS pressure sensors, as well as to digital infrasound sensors that output dynamic pressure values paired with timestamps.

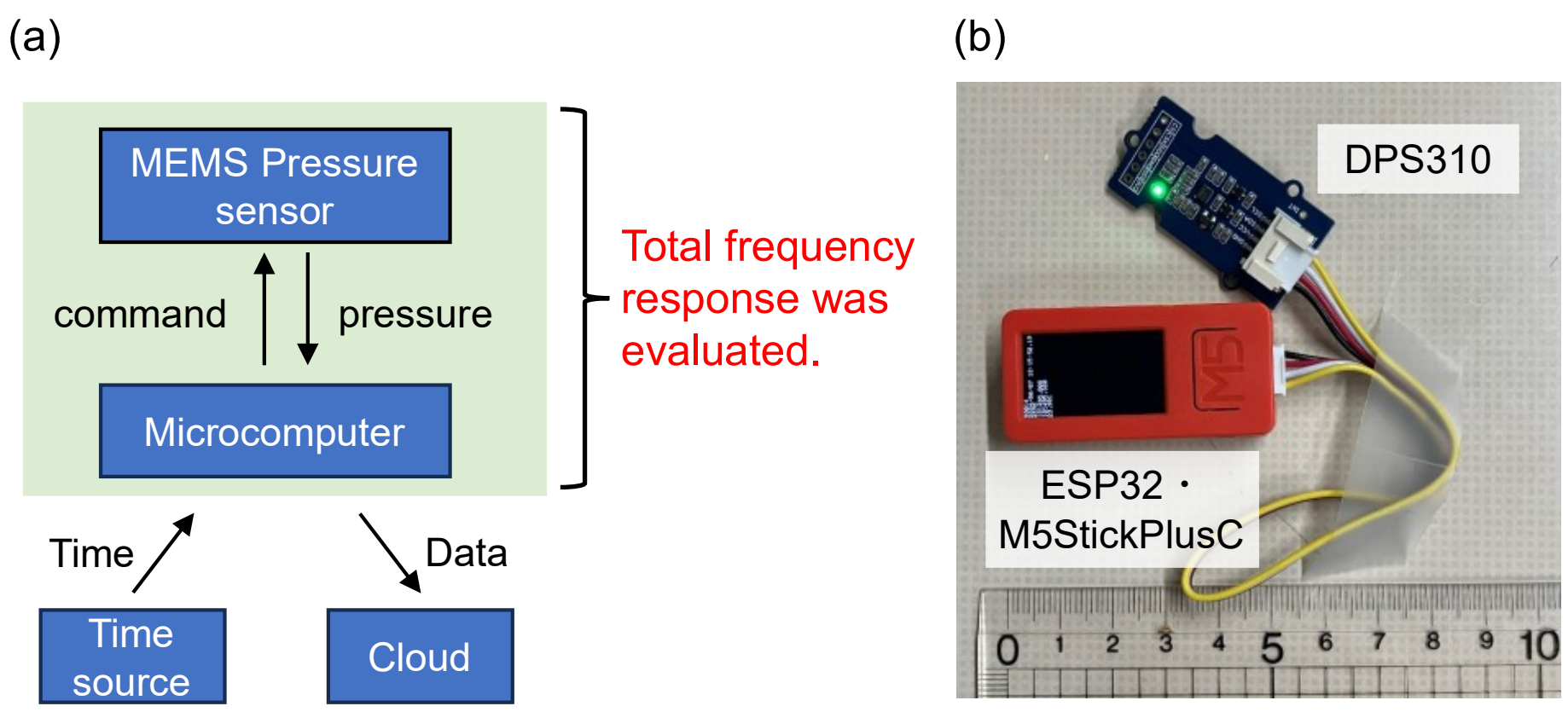

Figure 3. (a) Schematic of a MEMS module for infrasound measurement. (b) Photograph of the MEMS module evaluated in this study.

## 3.2 Signal processing

### 3.3.1 Sine fitting algorithm

The sensitivity modulus and phase were computed by approximating the digitized signal to a sinusoidal waveform using the least squares method, based on IEEE Standard 1057:2017 [29].

Let us denote the digitized sinusoidal signal recorded at time $t_1, \dots, t_\mathrm{n}$ as $\boldsymbol{y} = (y_1, \dots, y_\mathrm{n})^T$, where $(*)^T$ denotes the transpose of the vector or matrix. We assume that the data can be modelled using the following equation:

$$y_\mathrm{n}[\boldsymbol{\chi}] = A_0\cos(\omega t_\mathrm{n}) + B_0\sin(\omega t_\mathrm{n}) + C_0, \tag{1}$$

where $\boldsymbol{\chi} = (A_0, B_0, C_0)^T$ is a set of three unknown parameters and $\omega = 2\pi f$ is the angular frequency ($f$: frequency of interest). The sine fitting algorithm finds $\boldsymbol{\chi}$ which minimizes the sum of squared differences

$$\Gamma[\chi] = \frac{1}{N}\sum_{n=1}^{N}(y_\mathrm{n} - y_\mathrm{n}[\chi])^2, \tag{2}$$

where $y_\mathrm{n}$ is the experimental data and $y_\mathrm{n}[\boldsymbol{\chi}]$ is the modelled sine wave. In matrix notation, Equation (2) can be re-written as

$$\Gamma[\chi] = \frac{1}{N}(\boldsymbol{y} - \mathbf{D}(\omega)\boldsymbol{\chi})^T(\boldsymbol{y} - \mathbf{D}(\omega)\boldsymbol{\chi}), \tag{3}$$

where

$$\mathbf{D}(\omega) = \begin{bmatrix} \cos(\omega t_1) & \sin(\omega t_1) & 1 \\ \vdots & \vdots & \vdots \\ \cos(\omega t_{\mathrm{n}}) & \sin(\omega t_{\mathrm{n}}) & 1 \end{bmatrix}. \tag{4}$$

The least squares solution which minimizes Equation (4) is given by

$$\boldsymbol{\chi} = (\mathbf{D}^T\mathbf{D})^{-1}(\mathbf{D}^T\boldsymbol{y}), \tag{5}$$

where $(*)^{-1}$ denotes the inverse matrix. To convert $\boldsymbol{\chi}$ into the modulus $\hat{y}$ and phase $\varphi$ of the sine wave in Equation (5),

$$y_{\mathrm{n}} = \hat{y}\sin(\omega t_{\mathrm{n}} + \varphi) + C_0, \tag{6}$$

the following equations may be used:

$$\hat{y} = \sqrt{(A_0)^2 + (B_0)^2}, \tag{7}$$

$$\varphi = \arctan\left(\frac{B_0}{A_0}\right). \tag{8}$$

### 3.3.2 Sensitivity calculation

Let us denote the voltage data from the microphone as ($t_{\mathrm{a}}$, $v_{\mathrm{a}}$) and the pressure data from the DUT as ($t_{\mathrm{d}}$, $p_{\mathrm{d}}$), where $t_{\mathrm{a}}$ is the analog time, $v_{\mathrm{a}}$ is the voltage from the microphone, $t_{\mathrm{d}}$ is the digital time, and $p_{\mathrm{d}}$ is the pressure data from the MEMS module. The data length was set to $10/f$ ($f$: frequency of interest). As explained in Section 3.3.1, both signals were fitted with sine waves at $f$ to obtain the modulus $\hat{v}_{\mathrm{a}}$, $\hat{p}_{\mathrm{d}}$ and phase $\varphi_{\mathrm{a}}$, $\varphi_{\mathrm{d}}$:

$$v_{\mathrm{a}} = \hat{v}_{\mathrm{a}}\sin(2\pi f t_{\mathrm{a}} + \varphi_{\mathrm{a}}) + b_{\mathrm{a}}, \tag{9}$$

$$p_{\mathrm{d}} = \hat{p}_{\mathrm{d}}\sin(2\pi f t_{\mathrm{d}} + \varphi_{\mathrm{d}}) + b_d, \tag{10}$$

where $b_\mathrm{a}$ and $b_\mathrm{d}$ are offsets. When the sensitivity modulus and phase of the pre-calibrated microphone are $\hat{m}_\mathrm{mic}$ and $\varphi_\mathrm{mic}$, respectively, the modulus $\hat{p}_\mathrm{ref}$ and phase $\varphi_\mathrm{ref}$ of the sound pressure in the chamber can be calculated using the following equations:

$$\hat{p}_\mathrm{ref} = \frac{\hat{v}_\mathrm{a}}{\hat{m}_\mathrm{mic}}, \tag{11}$$

$$\varphi_\mathrm{ref} = \varphi_\mathrm{a} - \varphi_\mathrm{mic}. \tag{12}$$

Therefore, the sensitivity modulus ratio $s_\mathrm{d}$ and phase shift $\Delta\varphi_\mathrm{d}$ of the sound pressure measured by the MEMS module compared to the reference sound pressure are given by the following equations:

$$s_\mathrm{d} = \frac{\hat{p}_\mathrm{d}}{\hat{p}_\mathrm{ref}}, \tag{13}$$

$$\Delta\varphi_\mathrm{d} = \varphi_\mathrm{d} - \varphi_\mathrm{ref} - 2\pi f \Delta t_\mathrm{d-a}, \tag{14}$$

where $\Delta t_\mathrm{d-a}$ is the difference in the acquisition start time between the microphone and the MEMS module.

# 4 Results

Herein, we evaluated the specific MEMS module described in Section 3.2 as an example, and the results were discussed in conjunction with the analysis of uncertainty. The calibrated frequency responses are shown in Figure 4. Two modules of the same model were evaluated to examine unit-to-unit variability. For clarity, we refer to these two modules as Module1 and Module2, where Module1 consists of a MEMS pressure sensor1 and a Microcontroller1, and Module2 consists of a MEMS pressure sensor2 and a Microcontroller2.

Regarding the sensitivity modulus ratio $s_\mathrm{d}$ (hereafter 'modulus') of Module1, the average value agreed with the standard modulus (1.00) within 2%. Module2 exhibited slightly lower repeatability than Module1; however, its average value still matched the standard modulus within 5%. This variation seems to originate from the intrinsic self-noise of the MEMS pressure sensor. This suggests that, when conducting field measurement using these MEMS modules, the measured modulus is relatively reliable within the evaluated frequency range; however, variations owing to sensor-specific factors may still occur.

For phase shift $\Delta\varphi_\mathrm{d}$ (hereafter 'phase'), Module1 and Module2 showed a consistent time delay of 10 ms to 20 ms. This delay appears to reflect the time elapsed from the detection of sound pressure by the MEMS pressure sensor to the output of the timestamped value by the microcontroller. The results indicate that such time delays affect the phase measured in practical scenarios, and that considering them based on the calibration results could improve accuracy in source localization.

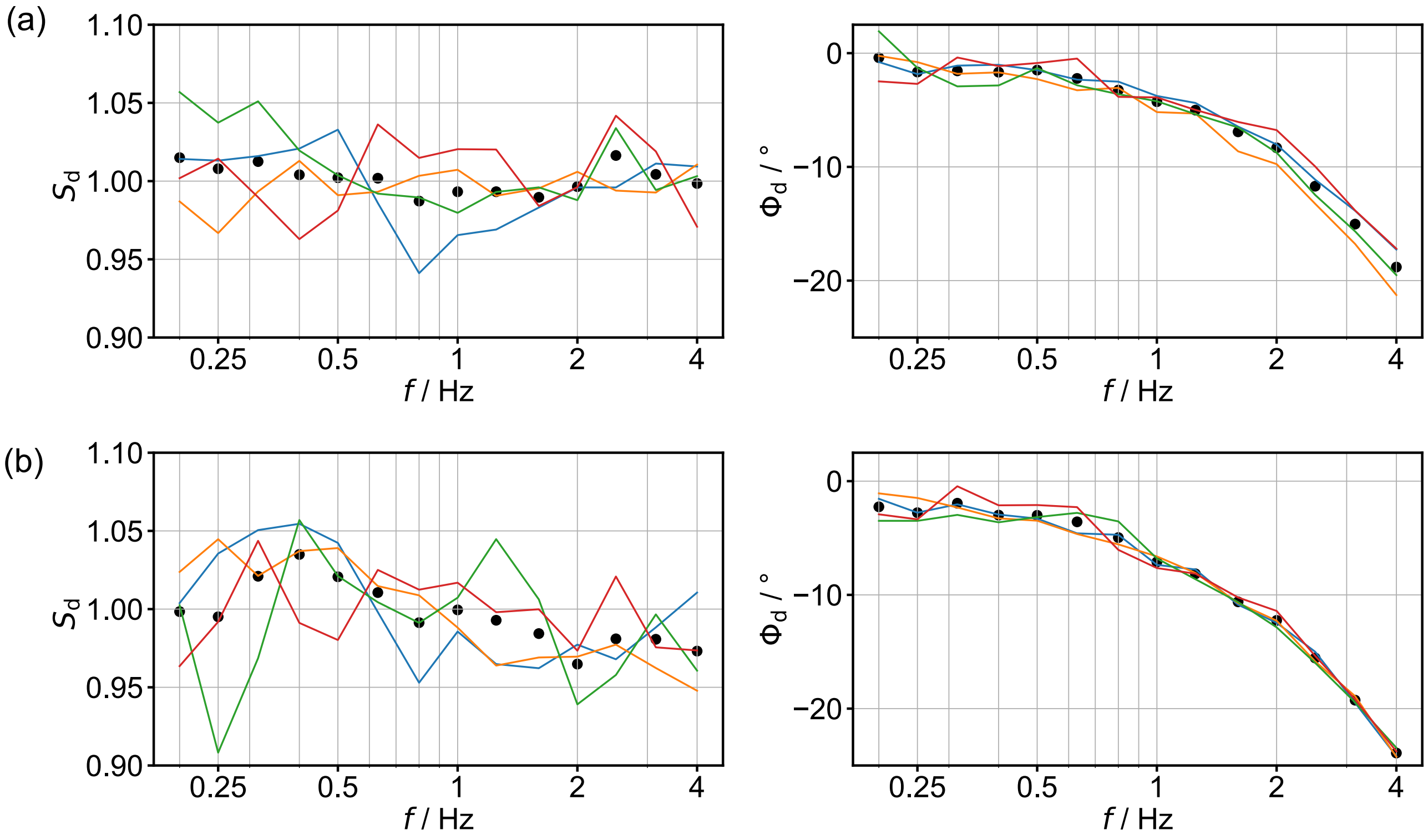

Figure 4. Sensitivity modulus and phase of (a) Module1 and (b) Module2. The black dots represent the average of four measurements. The thin colored lines represent the measured values for each of the four measurements.

# 5 Uncertainty discussion

## 5.1 Effect of clock inaccuracy on the sensitivity uncertainty

In the calibration of digital-output sensors, the MEMS modules rely on the system clock of their microcontroller to assign timestamps, whereas in conventional analog sensor calibration, the signals from both the DUT and the reference standard are digitized using the same A/D converter. Since the system clock in a microcontroller is less accurate than that of the clock in an A/D converter, its impact on the frequency response was evaluated.

To estimate the microcontroller's clock accuracy, phase calibration at $f = 1$ Hz was conducted 50 times at 3-minute intervals to observe long-term variations. If sufficient accuracy of the system clock was ensured, the phase lag would remain constant over time. However, the results in Figure 5 indicate that for Module1, a phase lead of approximately 4.1 milliseconds is added every 1000 seconds. This indicates that the system clock of the microcontroller in Module1 is delayed by 4.1 ppm. Note that these results were obtained in a well-controlled laboratory environment (temperature: (23 ± 1) °C, humidity: (50 ± 10)%), and that the accuracy of the system clock may be lower under in-field measurements.

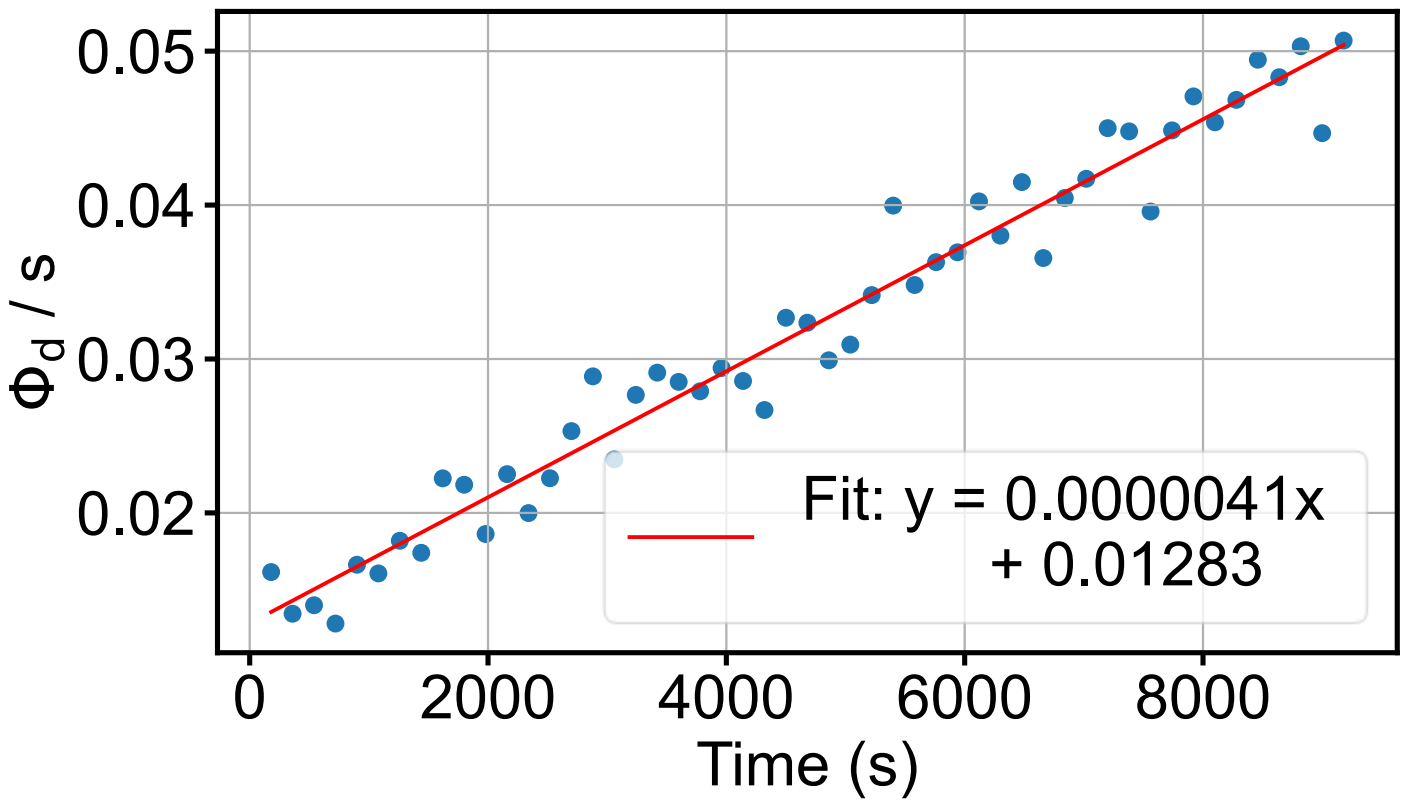

Figure 5. Long-term phase drift for Module1. Time correction using an external time-source was not applied during the measurement.

Then, the effect of this drift on the calibration was discussed from the following two viewpoints.

(1) Effect on the start time correction

The uncertainty in the system clock directly contributes to the uncertainty in $\Delta t_{\mathrm{d-a}}$ (the correction term for the start time difference between the microphone and the MEMS module) because the elapsed time recorded by the microcontroller's system clock was used to assign timestamps to pressure measurements. To minimize this effect, the microcontroller's system clock should be periodically reset. For example, in this experiment, it was reset every 10 minutes using the external PPS signal.

However, even when periodic resets are performed, the timestamping between resets still relies solely on the microcontroller's system clock, and some degree of drift is inevitable. Although the optimal solution would be to apply a correction for this time deviation, no correction had been previously applied because such clock drift had not been evaluated. Therefore, in the present study, we conservatively estimated the maximum possible uncertainty for Module1 by assuming that all measurements were taken exactly 10 minutes after the most recent reset:

$$u_{\Delta t_{\mathrm{d-a}}} = 600\ (\mathrm{s}) \times 4.1 \times 10^{-6} (\mathrm{ppm}) = 2.5 \times 10^{-3} (\mathrm{s}). \qquad (15)$$

This uncertainty can be reduced by implementing a correction mechanism that records the elapsed time since the last reset. Specifically, the timestamp deviation can be calculated and corrected by multiplying the recorded elapsed time by the slope of the regression curve in Figure 5. Nevertheless, even if this correction is applied, residual variations from the regression curve would remain, indicating the need for further detailed analysis for uncertainty evaluation.

(2) Wave distortion due to the drift of the system clock

Considering that the actual waveform is sampled not at $t_s$ but at $t_s(1+\Delta t)$ where $t_s$ denotes the nominal sampling period (s) and $\Delta t$ denotes the drift of the system clock (s/s), the effect of the system clock on the sine wave approximation was evaluated based on the process presented in Section 3.3.1. The effects of such drift on the calibration uncertainty—particularly in the case of vibration sensors—have been previously reported [30, 31].

When the data is acquired using an inaccurate sample clock, the obtained waveform can be expressed as follows:

$$y_n'[\boldsymbol{\chi}] = A_0\cos(\omega t_n') + B_0\sin(\omega t_n') + C_0, \tag{16}$$

where

$$t_n' = nt_s(1+\Delta t) \quad (n = 1, \dots, N), \tag{17}$$
$$N = \operatorname{int}\left(\frac{N_{cyc}}{ft_s}\right) \quad (N_{cyc}\text{: interger}).$$

Since the exact value of $\Delta t$ is unknown, sine fitting was performed under the assumption that the waveform was acquired at a fixed sampling period $t_s$. That is, the least squares solution $\boldsymbol{\chi}' = [A_0', B_0', C_0']^T$ was obtained using the following equation:

$$\chi' = (\mathbf{D}^T\mathbf{D})^{-1}(\mathbf{D}^T\boldsymbol{y}'). \tag{18}$$

where

$$\mathbf{D}(\omega) = \begin{bmatrix} \cos(\omega t_1) & \sin(\omega t_1) & 1 \\ \vdots & \vdots & \vdots \\ \cos(\omega t_n) & \sin(\omega t_n) & 1 \end{bmatrix},$$
$$t_n = nt_s \quad (n = 1, \dots, N).$$

The differences in sensitivity modulus and phase from sine wave approximation with and without drift were estimated under various conditions. As shown in Figure 6, the larger the drift and the longer the sample length, the greater effect on calibration results. In this experiment ( $\Delta t =$ 4.1 ppm (Module1), $N_{\mathrm{cyc}}$= 20, $t_{\mathrm{s}}$ = 70 ms), the effect of waveform distortion on the sensitivity was sufficiently smaller than those of other uncertainty components; thus, no correction was applied. However, in cases where clock drift is substantial or long-time measurements are conducted, the effect of waveform distortion on the sensitivity should be evaluated and corrected.

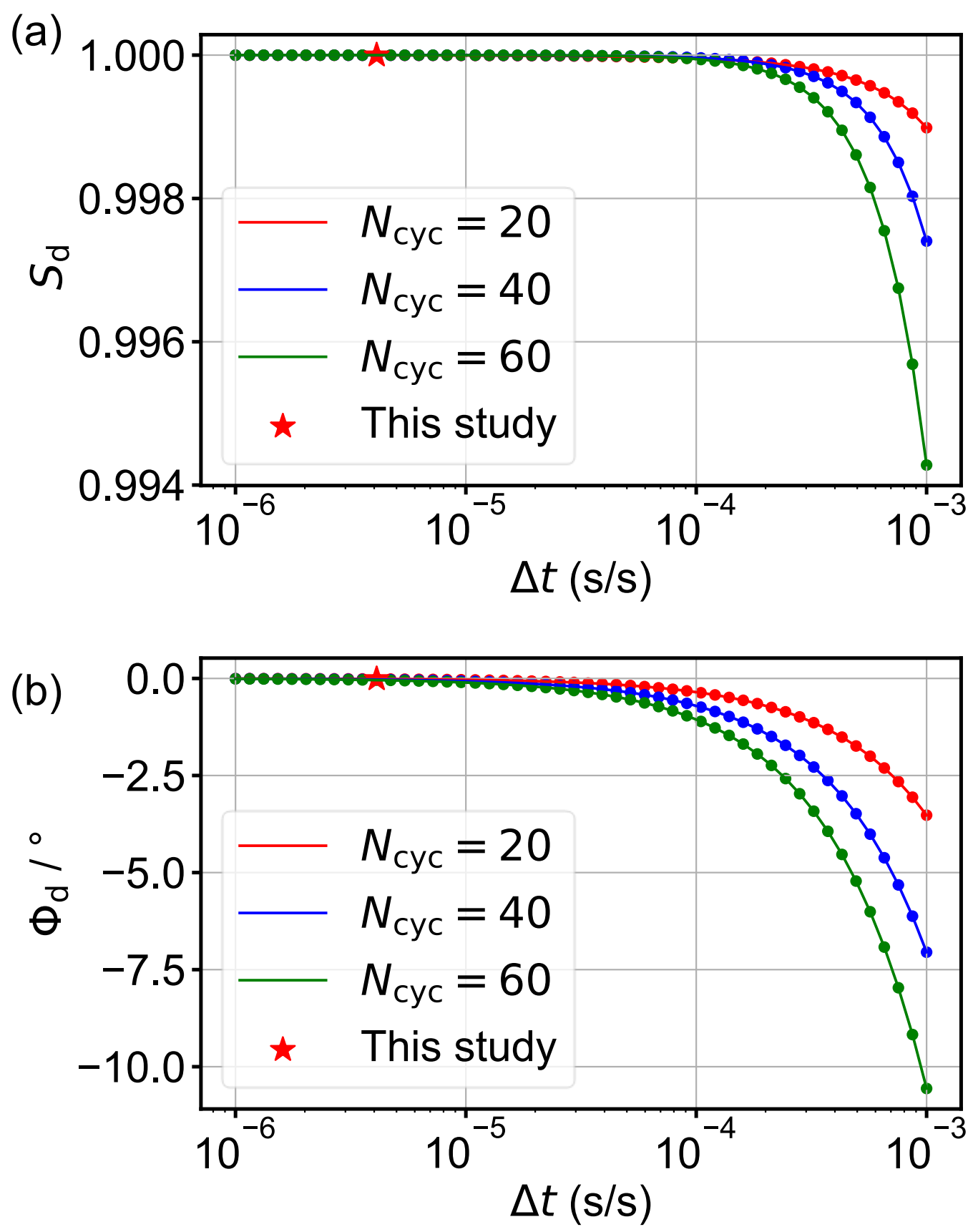

Figure 6. Dependency of the system clock drift on the (a) sensitivity modulus and (b) phase of a digital DUT sensor for various sample length. The star indicates the condition observed in Module1.

## 5.2 Uncertainty budgets

The uncertainty was estimated based on the model equation:

$$s_{\mathrm{d}} = \frac{\hat{p}_{\mathrm{d}}}{\frac{\hat{v}_{\mathrm{a}}}{\hat{m}_{\mathrm{mic}}}}, \tag{19}$$

$$\Delta\varphi_{\mathrm{d}} = \varphi_{\mathrm{d}} - \varphi_{\mathrm{a}} + \varphi_{\mathrm{mic}} - 2\pi f \Delta t_{\mathrm{d-a}}.$$

The uncertainty budgets of the sensitivity modulus and phase for Module1 are summarized in Table 1 and Table 2, respectively, and their frequency dependencies are shown in Figure 7. The expanded uncertainty, an interval which may be expected to encompass 95% distribution of values for $k = 2$, is 0.07 for the sensitivity modulus and 6.5° for the phase, respectively. Herein, the dominant source of uncertainty was repeatability, which was attributed to the lower accuracy of MEMS modules compared to precision analog sensors. However, below 1 Hz, the contribution of the uncertainty from the reference standard increased. This suggests the need to minimize the uncertainty in the primary calibration of the reference standard as much as possible. Considering the phase uncertainty, at frequencies above 1 Hz, the component 'microcomputer clock' became notable. To reduce this uncertainty, the microcontroller's system clock should be more frequently corrected using the external time-source.

The details of each uncertainty component are discussed below.

**(a)** Reference standard

The reference standard, BK Type 4160, was pre-calibrated by the laser pistonphone method. The associated uncertainty was adopted according to the previous study [28].

**(b)** Non-uniformity

To examine the uniformity of sound pressure within the chamber, calibrated microphones were placed at different locations in the chamber, and the modulus and phase differences in sound pressure between these points were measured. Below 20 Hz, the maximum differences were 0.6% for modulus and 0.1° for phase, respectively. We considered these maximum differences as Type B [32]; then, a factor of $1/\sqrt{3}$ was multiplied for the uncertainty budget.

**(c)** Digitizer gain

The uncertainty in the microphone voltage conversion by the digitizer was estimated as 0.35%, based on the specifications provided by the manufacturer.

**(d)** Repeatability

The standard deviation of the average value from four measurements was adopted. Note that the calculated standard deviation had an uncertainty of 42% because of the limited number of measurements [32] (Annex E.4).

**(e)** Digitizer asynchronization

The asynchronization between two digitizers, PXIe-4461 and PXIe-5172, was examined by comparing their timing differences when injecting the same voltage from one source. The maximum difference was assessed as 10 µs. Assuming a rectangular distribution, a factor of $^{1}/_{\sqrt{3}}$ was multiplied for the uncertainty budget.

**(f)** Microcomputer clock

The maximum uncertainty caused by the drift of microcomputer's system clock was estimated as 2.5 ms (see Section 4.2). Assuming a rectangular distribution, a factor of $^{1}/_{\sqrt{3}}$ was multiplied for the uncertainty budget.

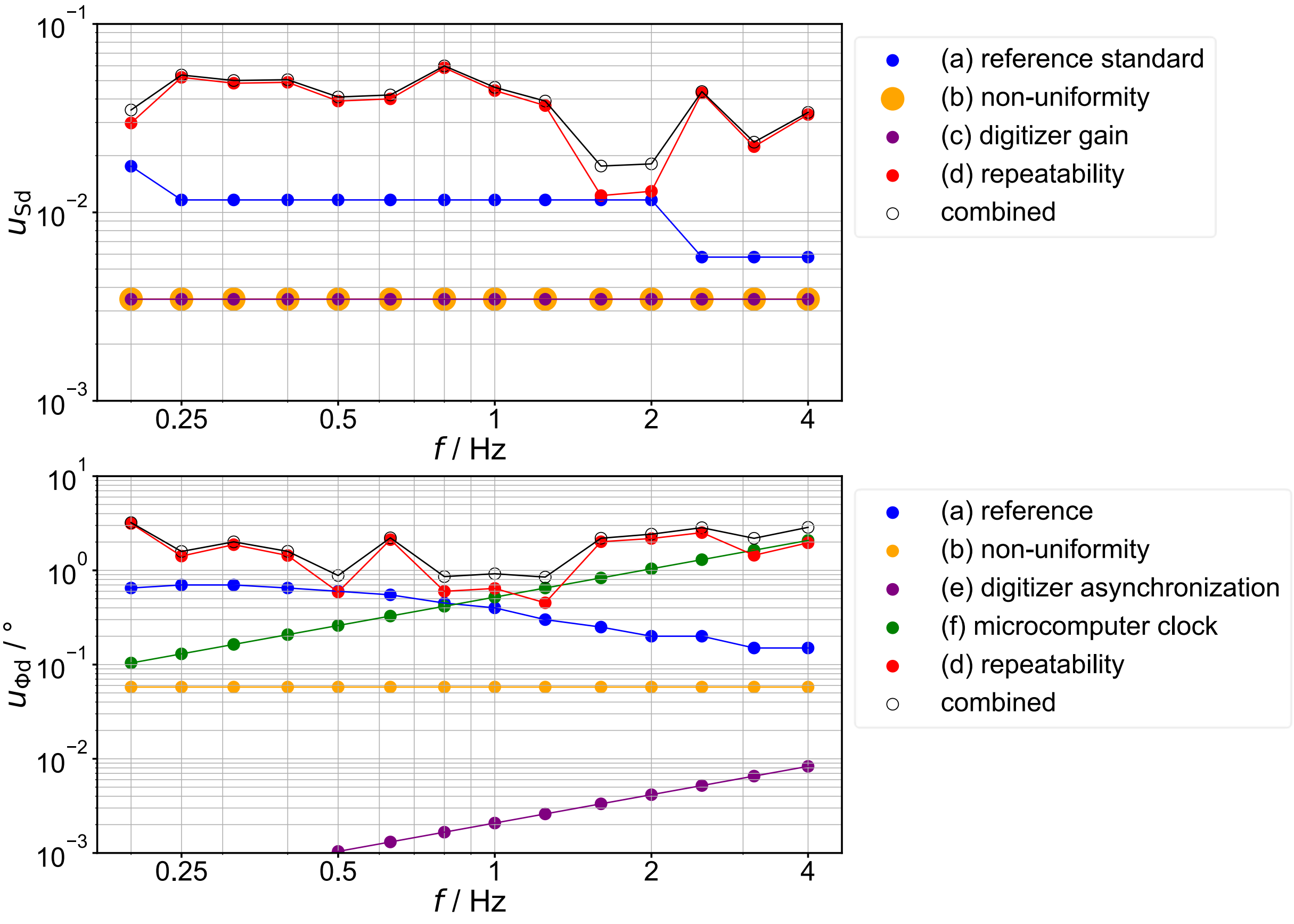

Figure 7. Uncertainty components of the sensitivity modulus and phase of Module1.

Table 1 Uncertainty budget for the sensitivity modulus of Module1

| Symbol | Component | Type | Distribution | Uncertainty value at measurement frequency | |
|---|---|---|---|---|---|
| | | | | 0.2 Hz | 4 Hz |
| $u_{(\hat{m}_{mic})}$ | (a) Reference standard | A [1)] | Normal | 0.0176 | 0.0058 |
| $u_{(\hat{p}_d/\hat{v}_a)}$ | (b) Non-uniformity | B [2)] | Rectangular | 0.0035 | 0.0035 |
| $u_{(\hat{v}_a)}$ | (c) Digitizer gain | B | Rectangular | 0.0035 | 0.0035 |
| $u_{(R)}$ | (d) Repeatability | A | Normal | 0.0298 | 0.0331 |
| $u_c$ | Combined standard uncertainty | | | 0.0350 | 0.0339 |
| $U$ | Expanded uncertainty ($k = 2$) [3)] | | | 0.0699 | 0.0679 |

Table 2 Uncertainty budget for the sensitivity phase (°) of Module1

| Symbol | Component | Type | Distribution | Uncertainty value at measurement frequency | |
|---|---|---|---|---|---|
| | | | | 0.2 Hz | 4 Hz |
| $u_{(\varphi_{mic})}$ | (a) Reference standard | A | Normal | 0.650 | 0.150 |
| $u_{(\varphi_d-\varphi_a)}$ | (b) Non-uniformity | B | Rectangular | 0.058 | 0.058 |
| $u_{(\varphi_a)}$ | (e) Digitizer asynchronization | B | Rectangular | 0.000 | 0.008 |
| $u_{(\Delta t_{d-a.})}$ | (f) Microcomputer clock | B | Rectangular | 0.104 | 2.078 |
| $u_{(R)}$ | (d) Repeatability | A | Normal | 3.144 | 1.962 |
| $u_c$ | Combined standard uncertainty | | | 3.212 | 2.863 |
| $U$ | Expanded uncertainty ($k = 2$) [3)] | | | 6.425 | 5.725 |

1) Type A evaluation: Uncertainty evaluation from repeated observations under the same measurement conditions [32].

2) Type B evaluation: Uncertainty evaluation by scientific judgement based on all of the available information on the possible variability of input quantity [32].

3) For some frequencies, the component “Repeatability” notably contributes to the combined uncertainty. As a result, the total effective degrees of freedom obtained from the Welsh–Satterhwaite formula is below 10, meaning that the coverage factor $k$ should be 3 or more to ensure that the measurement results are encompassed within $y \pm ku_c$ interval with a probability of 95 % ($y$: the best estimate of the value attributable to the measurand). To allow $k = 2$ as a coverage factor, we used the Bayesian uncertainty by multiplying $\sqrt{{m-1}/{m-3}}$ with the repeatability uncertainty ($m$: number of the repeatability measurements, in this case 4) [33].

## 6 Conclusion

In this study, we developed a comparison calibration system specifically for digital-output infrasound sensors. First, the general requirements for the calibration system were defined and organized; then, the actual system was constructed accordingly. As a case study, the sensitivity modulus and phase of MEMS modules composed of a MEMS pressure sensor DPS310 and a microcomputer ESP32 were evaluated in the frequency range of 0.2 Hz to 4 Hz. Uncertainty analysis revealed that the main sources of uncertainty were repeatability and the inaccuracy of the DUT clock, rather than the inherent uncertainty of the system itself. This indicates that the developed system is not only applicable to the specific MEMS modules evaluated herein but also to various types of digital-output infrasound sensors with higher accuracy and/or lower self-noise.

## Appendix. Timestamp acquisition system of the MEMS module

Here, the details on obtaining time information with the microcomputer ESP32 are provided. Generally, microcomputers acquire time information from external time-sources such as the Global Positioning System (GPS) or an NTP server. In this experiment, we adopted a method that is considered the most precise for obtaining time information, simulating the use of both NTP and GPS; specifically, time data with second-level precision were acquired via communication with an NTP server, whereas sub-second time data were determined using the system clock of ESP32 corrected by the PPS signal output from time-sources such as GPS or time distributor (Figure S1(a)). Without correction using the PPS signal, network delays between the NTP server and the ESP32 could induce timing errors on the order of 1 second.

Figure S1(b) depicts the scheme employed for acquiring sub-second time data. The ESP32, equipped with a system clock that measures the elapsed time in milliseconds since powering on, detects the rising edge of the PPS signal and records the elapsed time $m_{\mathrm{c}}$ (ms) just before the measurement. By calculating the difference between elapsed time $m_{\mathrm{t}}$ (ms) when pressure data is received from DPS310 and $m_{\mathrm{c}}$ (ms), the absolute sub-second time can be determined with millisecond precision.

The specific implementation was as follows. Just before the start of measurement, the time information with second-level precision was obtained from the NTP server (ntp.nict.jp) managed by the National Institute of Information and Communications Technology (NICT). Simultaneously, the elapsed time $m_{\mathrm{c}}$ (ms) corresponding to the rising edge of the PPS signal output by a time distribution device was recorded, and this moment was defined as ($\mathrm{hh_{start}}$. $\mathrm{mm_{start}}$.$\mathrm{ss_{start}}$.000). Subsequently, the timestamp for pressure measurements, ( $\mathrm{hh_{d}}$ . $\mathrm{mm_{d}}$ . $\mathrm{ss_{d}}$ .$\mathrm{SSS_{d}}$), was determined by adding $m_{\mathrm{t}}-m_{\mathrm{c}}$ (ms) to ( $\mathrm{hh_{start}}$ . $\mathrm{mm_{start}}$.$\mathrm{ss_{start}}$.000).

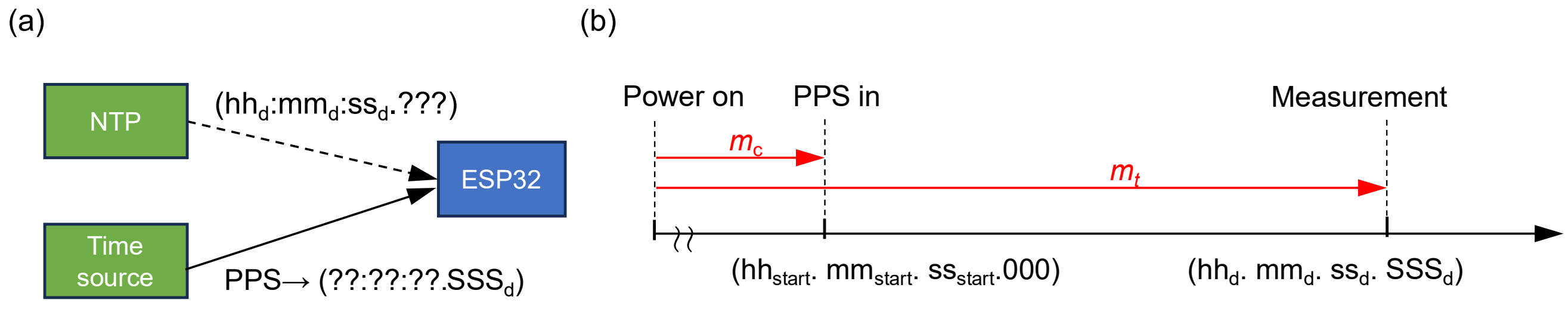

Figure S1. Methods for obtaining timestamps on the ESP32: (a) through communication with an NTP server and by obtaining a PPS signal from an external time-source. (b) Detailed scheme for obtaining accurate time by combining PPS signal and the system clock of ESP32.

## Data availability statement

The data cannot be made publicly available upon publication because they are not available in a format sufficiently accessible or reusable by other researchers. The data that support the findings of this study are available upon reasonable request from the authors.

## Acknowledgements

The authors acknowledge Prof. Kensuke Nakajima (Kyusyu University) for providing us with the original program to control the MEMS modules. The MEMS modules in this study were provided by Prof. Masa-Yuki Yamamoto (Kochi University of Technology). This work was partly supported by the commissioned research (No. 22605) by National Institute of Information and Communications Technology (NICT), Japan.